\documentstyle[11pt,a4]{article}
\newcommand{\CMP}[1]{{\em Commun. Math. Phys.} {\bf {#1}}}
\newcommand{\JMP}[1]{{\em J.~Math. Phys.} {\bf {#1}}}

\newcommand{\Ann}[1]{{\em Ann. Phys.} {\bf {#1}}}
\newcommand{\xx}{\!\stackrel {\scriptscriptstyle \times}%
{\scriptscriptstyle \times}\!}
\newcommand{\dLam}{\Lambda^*}
\newcommand{\dLamF}{\Lambda^*_{odd}}
\newcommand{\dInt}{\hat{\int}_{\dLam}}
\newcommand{\dIntF}{\hat{\int}_{\dLamF}}
\newcommand{\dIntp}{\hat{\int}_{\dLam\backslash \{0\}}}
\newcommand{\Lam}{\Lambda}
\newcommand{\Int}{\int_{\Lam}}
\newcommand{\ddel}{\hat{\delta}}
\newcommand{\Del}{\hat{\rm d}}

\newcommand{\U}{{\rm U}}
\newcommand{\tto}{\stackrel{R}{\to}}
\newcommand{\half}{\mbox{$\f{1}{2}$}}

\newcommand{\eq}{\begin{equation}}
\newcommand{\eqend}{\end{equation}}
\newcommand{\eqa}{\begin{eqnarray}}
\newcommand{\nonueqa}{\begin{eqnarray*}}
\newcommand{\eqaend}{\end{eqnarray}}
\newcommand{\nonueqaend}{\end{eqnarray*}}
\newcommand{\nonu}{\nonumber \\ \nopagebreak}
\newcommand{\bma}[1]{\begin{array}{#1}}
\newcommand{\ema}{\end{array}}
\newcommand{\bc}{\begin{center}}
\newcommand{\ec}{\end{center}}

\newcommand{\Ref}[1]{(\ref{#1})}

\newcommand{\dd}{{\rm d}}
\newcommand{\ee}[1]{\mbox{{\rm e}}^{#1}}
\newcommand{\ii}{{\rm i}}
\newcommand{\OO}{{\cal O}}

\renewcommand{\vec}[1]{\mbox{\boldmath ${#1}$}}
\newcommand{\vx}{\vec{x}}

\newcommand{\om}{\omega}

\newcommand{\lam}{\lambda}
\renewcommand{\phi}{\varphi}
\newcommand{\del}{\delta}
\newcommand{\Om}{\Omega}

\newcommand{\tet}{\theta}

\newcommand{\eps}{\varepsilon}

\newcommand{\al}{\alpha}
\newcommand{\ga}{\gamma}
\newcommand{\la}{\lambda}
\newcommand{\si}{\sigma}
\newcommand{\De}{\Delta}

\newfam\msbfam
\batchmode\font\twelvemsb=msbm10 scaled\magstep1 \errorstopmode
\ifx\twelvemsb\nullfont\def\Bbb{\bf}
\else   \catcode`\@=11
        \font\tenmsb=msbm10 \font\sevenmsb=msbm7 \font\fivemsb=msbm5
        \textfont\msbfam=\tenmsb
        \scriptfont\msbfam=\sevenmsb \scriptscriptfont\msbfam=\fivemsb
        \def\Bbb{\relax\ifmmode\expandafter\Bbb@\else
                \expandafter\nonmatherr@\expandafter\Bbb\fi}
        \def\Bbb@#1{{\Bbb@@{#1}}}
        \def\Bbb@@#1{\fam\msbfam\relax#1}
        \catcode`\@=\active
\fi
\newcommand{\R}{{\Bbb R}}
\newcommand{\C}{{\Bbb C}}
\newcommand{\Z}{{\Bbb Z}}
\newcommand{\N}{{\Bbb N}}

\newcommand{\f}{\frac}

\newcommand{\cD}{{\cal D}}

\newcommand{\cH}{{\cal H}}

\newcommand{\cU}{{\cal U}}
\newcommand{\cS}{{\cal S}}

\newcommand{\ccr}[2]{{[} {#1},{#2} {]} }
\newcommand{\car}[2]{{\{} {#1},{#2} {\}} }

\newcommand{\normal}[1]{:{#1}:\;}

\newcounter{saveeqn}
\newcounter{App} 
\newcommand{\app}{%
\stepcounter{App}%
\setcounter{saveeqn}{\value{equation}}%
\setcounter{equation}{0}%
\renewcommand{\theequation}{\Alph{App}\arabic{equation}} }
\newcommand{\appende}{%
\setcounter{equation}{\value{saveeqn}}%
\renewcommand{\theequation}{\arabic{equation}}  }
\newcommand{\alpheqn}{%
\stepcounter{equation}
\setcounter{saveeqn}{\value{equation}}%
\setcounter{equation}{0}%
\renewcommand{\theequation}{\arabic{saveeqn}\alph{equation}} }
\newcommand{\reseteqn}{\setcounter{equation}{\value{saveeqn}}%
\renewcommand{\theequation}{\arabic{equation}} }
\newcounter{asaveeqn}
\newcommand{\aalpheqn}{%
\stepcounter{equation}
\setcounter{asaveeqn}{\value{equation}}%
\setcounter{equation}{0}%
\renewcommand{\theequation}{\Alph{App}\arabic{asaveeqn}\alph{equation}} }
\newcommand{\areseteqn}{\setcounter{equation}{\value{asaveeqn}}%
\renewcommand{\theequation}{\Alph{App}\arabic{equation}} }
\begin{document}

\begin{flushright}
UWThPh-1996-5\\
August 30, 1996
\end{flushright}
\vspace{1.5cm}
\renewcommand{\thefootnote}{\alph{footnote}}
\begin{center}

{\Large \bf The Luttinger--Schwinger Model}\\
\vspace{1.5cm}
{\large Harald Grosse$^*$, Edwin Langmann$^{**}$ and
Ernst Raschhofer$^{*,}$\footnote{Supported in part by the  ``Fonds
zur F\"orderung von Wissenschaft und Forschung in \"Osterreich'' under
contract P8916-PHY}} \\
\end{center}
\vspace*{0.5cm}
$^*${\em Institut f\"ur Theoretische Physik,
Universit\"at Wien, A-1090 Wien, Austria} \\
$\;^{**}${\em Theoretical Physics, Royal Institute of Technology, S-10044
Stockholm, Sweden}\\
\vspace{2cm}

\setcounter{footnote}{0}
\renewcommand{\thefootnote}{\arabic{footnote}}

\begin{abstract}
We study the Luttinger--Schwinger model, i.e.\ the (1+1) dimensional model
of massless Dirac fermions with a non-local 4--point interaction coupled to
a $\U(1)$-gauge field. We work within the Hamiltonian framework on
the cylinder, and construct the field operators and observables
as well--defined operators on the physical Hilbert space.  The complete
solution of the model is found using the boson--fermion correspondence, and
the formalism for calculating all gauge invariant Green functions is
provided.  We discuss the role of anomalies and show how the existence of
large gauge transformations implies a fermion condensate in all physical
states.  The meaning of regularization and renormalization in our
well--defined Hilbert space setting is discussed.  We illustrate the latter by
performing the limit to the Thirring--Schwinger model where the
interaction becomes local.
\end{abstract}
\newpage

\section{Introduction}

Since the early days of quantum field theory (QFT) 1+1 dimensional models
have attracted much attention. They have been extremely valuable to
develop general ideas and intuition about the structure of QFT.
The eldest and perhaps most popular of these 1+1 D models bear the names of
Thirring \cite{Thirring} -- Dirac fermions interacting with a local
current-current interaction -- and Schwinger \cite{Schwinger} -- quantum
electrodynamics with fermions. The models originated
in particle physics and
therefore, in order to have Lorentz invariance, were considered mainly on
infinite space $\R$ (see e.g.\ \cite{CHu,CW,CRW} and references therein).
Then one has to deal with infrared (infinite space volume) divergences in
addition to singularities coming from the ultraviolet (short distances).
In case the fermions are massless, both
models are soluble \cite{Schwinger,Klaiber,swieca} and a very detailed
picture of their properties can be obtained.  Another related model
originated from solid state physics and is due to Luttinger
\cite{Luttinger} -- massless Dirac fermions on spacetime $S^1\times\R$
interacting with a non-local current-current interaction 
(Lorentz invariance is nothing natural to ask for in solid
state physics).
The Luttinger
model shows that an interacting fermion system in one space dimension
need not behave qualitatively similar to free fermions 
but rather has properties similar to a boson system.  Such behaviour 
is generic for 1+1 D interacting fermion models and is denoted as Luttinger
liquid in solid state physics, in contrast to Landau liquids common in 3+1 D.

To consider the Luttinger model on compact space has the enormous technical 
advantage that infra red
(IR) problems are absent, and one can concentrate on the
short distance (UV) properties which are rather simple due
to the non-locality of the interaction.  In fact, this allows 
a construction of the interacting model on the Fock space of {\em free}
fermions \cite{MattisLieb,HSU,CH}
and one directly can make
use of mathematical results from the representation theory of the affine
Kac-Moody algebras.  Such an approach was recently given for QCD with
massless fermions \cite{LS3}.
As shown by Manton \cite{Manton}, the Schwinger model on compact space $S^1$ 
allows a
complete understanding of the UV divergences and anomalies and their
intriguing interplay with gauge invariance and vacuum structure.

In the present paper we study the extension of the Luttinger model
obtained by coupling it to a dynamical electromagnetic field.  For
vanishing Luttinger (4-point) interaction our model therefore reduces to
the Schwinger model as studied by Manton \cite{Manton}, and for vanishing
electric charge to the Luttinger model \cite{MattisLieb}.
Since our approach is in Minkowski space and provides a
direct construction of the field-- and observable algebras of the model on
a physical Hilbert space, it is conceptually quite different from the path
integral approach, and we believe it adds to the physical understanding of
these models.

The plan of the paper is as follows.  In Section 2 the construction of
the model is given.  To fix notation, we first summarize the classical
Hamiltonian formalism.  We then construct the physical Hilbert
space and discuss the
non-trivial implications of anomalies (Schwinger terms) and gauge
invariance.  In Section 3 the model is solved by bosonization, and a
method for calculating all Green functions is explained.  As an example the
equal time 2-point functions are given.  In Section 4 we comment on
regularization and renormalization in our setting.  We discuss the limit to
the Thirring-Schwinger model where the 4-point interaction becomes local
and space infinite.  We end with a short summary in Section 5.  A
summary of the mathematical results needed and some details of calculations
are deferred to the appendix.

\section{Constructing the model}
\subsection{Notation}

Spacetime is the cylinder with $x=x^1\in \Lam\equiv
[-L/2,L/2]$ the spatial coordinate and $t=x^0\in\R$ time. We have one
Dirac Fermion field $\psi_{\sigma}(\vx)$ and one Photon field
$A_\nu(\vx)$ (here and in the following,
$\sigma,\sigma'\in\{+,-\}$ are spin indices,
$\mu,\nu\in\{0,1\}$ are spacetime indices, and
$\vx=(t,x),\vec{y}=(t',y)$ are spacetime
arguments).

The action defining the Luttinger-Schwinger model
is\footnote{unless otherwise stated,
repeated indices are summed over throughout the paper}$^,$
\footnote{$\partial_\nu\equiv\partial/\partial x^\nu$;
our metric tensor is $g_{\mu\nu}=diag(1,-1)$}
\eqa
\label{1}
\cS = \int d^2\vec{x}\left(-\f{1}{4}F_{\mu\nu}(\vec x)F^{\mu\nu}(\vec x) +
\bar\psi(\vec x) \gamma^\nu\left(-\ii\partial_\nu +
e A_\nu(\vec x)\right)\psi(\vec x)\right) \nonu
- \int d^2\vec{x}\int d^2\vec{y}\, j_\mu (\vec x)v(\vec x
-\vec y) j^\mu(\vec y)
\eqaend
where $F_{\mu\nu}=\partial_\mu A_\nu-\partial_\nu
A_\mu$
and $\gamma^\nu\equiv (\gamma^\nu)_{\sigma\sigma'}$ are Dirac matrices
which we take as $\gamma^0=\sigma_1$ and
$\gamma^1=\ii\sigma_2$, and $\gamma_5=-\gamma^0\gamma^1=\sigma_3$
($\sigma_i$ are Pauli spin matrices).
As usual, the fermion currents are
$
j_\nu = \bar\psi\gamma_\nu\psi,
$
and we assume the 4--point interaction to be instantaneous (local in time)
\eq
v(\vec x-\vec y) = \delta(t-t')V(x-y)
\eqend
where the interaction potential is parity
invariant, $V(x)=V(-x)$.
As in case of the Luttinger  model \cite{HSU} we will also have to assume
that this potential is not `too strong', or more precisely that the Fourier
coefficients
\alpheqn
\eq
\label{condition}
W_k = \f{1}{8\pi}\Int\dd{x}\, V(x)\ee{-\ii kx} = W_{-k}= W_k^*,\quad
k\f{L}{2\pi} \in\Z
\eqend
of the potential obey the conditions
\eq
\label{condition1}
-1-\f{e^2}{\pi k^2} < W_k < 1\quad\forall k\quad
\mbox{ and }\quad \sum_k |kW_k^2|<\infty .
\eqend
\reseteqn

{}From the action \Ref{1} we obtain the canonical momenta $\Pi_{A_0(x)}
\simeq 0$, $\Pi_{A_1(x)} = F_{01}(x) = E(x)$ etc.\
(here and in the following, we set $t=0$ and make
explicit the dependence on the spatial coordinate only) resulting in the
Hamiltonian ($\psi^*\equiv \bar\psi\gamma^0$)
\eqa
\label{4}
H=\Int \dd{x} \left( \f{1}{2} E(x)^2 +
\psi^*(x)\gamma_5\left(-\ii\partial_1 + e A_1(x)\right)
\psi(x)\right) +  4\Int \dd{x} \dd{y} \,  \rho^+(x)V(x-y)\rho^-(y),
\eqaend
and the Gauss' law
\eq
\label{5}
G(x) = -\partial_1 E(x) + e \rho(x) \simeq 0 \, .
\eqend
We introduced chiral fermion currents
\eq
\rho^\pm(x) = \psi^*(x)\f{1}{2}(1\pm\gamma_5)\psi(x) \label{chiralcu}
\eqend
so that fermion charge-- and momentum density $\rho=j^0$ and $j=j^1$
can be written as
\eqa
\rho(x) &=& \rho^+(x) + \rho^-(x)\nonu
j(x) &=& \rho^+(x) - \rho^-(x).
\eqaend

\subsection{Observables}
\label{obs}
The observables of the model are all gauge invariant operators.
They leave invariant physical states. The ground state
expectation values of these operators are the Green functions we are interested
in. For later reference we write down the action of static gauge
transformations i.e.\ differentiable maps $\Lam\to\U(1), x\mapsto \ee{\ii
\alpha(x)}$,
\eqa
\label{gaugetrafo}
\psi_\sigma(x)&\to& \ee{\ii \alpha(x)}\psi_\sigma(x) \nonu
A_1(x) &\to& A_1(x) - \f{1}{e}\f{\partial \alpha(x)}{\partial x} \\
E(x) &\to& E(x)\, . \nonumber
\eqaend
These obviously leave our Hamiltonian and Gauss' law invariant.  We note
that every gauge transformation can be decomposed into a {\em small} and a
{\em large} gauge transformation,
$
\alpha(x)=\alpha_{small}(x) + \alpha_{large}(x),
$
where
\eq
\alpha_{large}(x) = n\f{2\pi x}{L} \quad (n\in\Z), \quad
\alpha_{small}\left(-\f{L}{2}\right)=
\alpha_{small}\left(\f{L}{2}\right)
\eqend
with $n=\f{\alpha(L/2)-\alpha(-L/2)}{2\pi L}$.  The large gauge
transformations correspond to $\Pi_1(S^1)=\Z$ and play an important role in
the following, as expected from general arguments \cite{Jackiw}.
It is important to note that Gauss' law \Ref{5} requires physical states
only to be invariant under small (but {\em not} under large) gauge
transformations.

All gauge invariant objects which one can construct 
from $A_1(x)$ (at fixed time) are functions of
\eq
Y= \f{1}{2\pi} \int_{\Lam}\dd y A_1(y) \, .
\eqend
In fact, $Y$ above is only invariant with respect to small gauge
transformations and changes by multiples of $1/ e$ under the large ones.
Thus the quantity which is invariant under all gauge transformations is
$\ee{\ii 2\pi e Y}$ which is equal to the Wilson line (holonomy)
\eq
\label{Wilson}
W[A_1] = \ee{\ii e\int_{\Lam}\dd y A_1(y)} \: .
\eqend
The fermion fields are not gauge invariant, but by attaching
parallel transporters to them one obtains field operators
\eq
\label{chi}
\chi_\sigma(x) = \ee{\ii e\int_r^x\dd y  A_1(y) }\psi_\sigma(x)\, ,
\quad r\in\Lam
\eqend
which obviously are invariant under all (small and large) gauge
transformations \Ref{gaugetrafo} with $\alpha(r)=0$; $r$ is a spatial
point which we can choose arbitrarily.  Note that these fields also obey
CAR but are {\em not} antiperiodic: they obey $\chi_\sigma(L/2)=
-W[A_1]\chi_\sigma(-L/2)$ where $W[A_1]$ is the Wilson line above.
Bilinears of these operators are the meson operators
$$
M_{\sigma\sigma'}(x,y) = \chi^*_\sigma(x)\chi_{\sigma'}(y) \, .
$$
These are invariant under all static gauge
transformations and thus can be used as building blocks of the Green
functions we are interested in.

\subsection{The quantum model}
In the following we find it convenient to work in Fourier space.
We introduce the following useful notation. Fourier space for even
(periodic) functions is
\alpheqn
\eq
\label{a}
\dLam\equiv \left\{\left. k=\f{2\pi}{L} n \right| n\in\Z\right\}\quad \, .
\eqend
As we use fermions with odd (anti--periodic) boundary conditions we
also need
\eq
\dLamF\equiv \left\{\left. k=\f{2\pi}{L} \left(n+\f{1}{2}\right)
\right| n\in\Z\right\}.
\eqend
{}For functions $\hat f$ on Fourier space we write
\eq
\dInt \Del k \hat f(k) \equiv \sum_{k\in\dLam} \f{2\pi}{L} \hat f(k)
\eqend
\reseteqn
and similarly for $\dLamF$ (we will use the same symbols $\delta$ and
$\hat\delta$ also in the latter case). Then the appropriate
$\del$-function satisfying $\dInt\Del
q\, \ddel(k-q)\hat f(q)=\hat f(k)$ is
$\ddel(k-q) \equiv \f{L}{2\pi} \del_{k,q}$.

{}For the Fourier transformed operators we use the following conventions,
\alpheqn
\eq
\label{10a}
\hat\psi^{}_{\sigma}(q) =
\Int \f{\dd{x}}{\sqrt{2\pi}} \psi^{}_{\sigma}(x)
\ee{-\ii qx},\quad
\hat\psi^{*}_{\sigma}(q)=\hat\psi^{}_{\sigma}(q)^*
\quad (q\in\dLamF)
\eqend
(as mentioned, we use anti--periodic boundary conditions for the fermions),
\eq
\label{A1}
\hat A_1(k) = \Int \f{\dd{x}}{2\pi}
A_1(x) \ee{-\ii kx} \quad (k\in\dLam)
\eqend
and in the other cases
\eq
\label{other}
\hat Y(k) = \Int \dd{x}\,
Y(x) \ee{-\ii kx}\quad (k\in\dLam)\quad \mbox{ for $Y=E,\rho^{\pm},\rho,j,V$}
\eqend
\reseteqn
Following \cite{HSU} we also find it convenient to introduce $W_k=\hat
V(k)/8\pi$ (cf.\ \Ref{condition}).  With that the non-trivial C(A)CR in
Fourier space are
\eqa
\label{fcacr}
\ccr{\hat A_1(p)}{\hat E(k)} &=& \ii\ddel(k+p) \nonu
\car{\hat\psi_{\sigma}(q)}{\hat\psi^*_{\sigma'}(q')}
&=& \delta_{\sigma\sigma'}\ddel(q-q') .
\eqaend

The essential physical requirement determining the construction of the
model and implying a non-trivial quantum structure is positivity of
the Hamiltonian on the physical Hilbert space.
It is well-known that it forces one to use a
non-trivial representation of the field operators of the model. The
essential simplification in (1+1) (and not possible in higher)
dimensions is that one can use a quasi-free representation for the
fermion field operators corresponding to ``filling up the Dirac sea''
associated with the {\em free} fermion Hamiltonian, and for the
photon operators one can use a naive boson representation.
This will be verified for our model
for the class of potentials $V$ obeying
(\ref{condition},b).

So the full Hilbert space of the model is $\cH = \cH_{\rm
Photon}\otimes \cH_{\rm Fermion}$.
For $\cH_{\rm Photon}$ we take the boson Fock space
generated by boson field operators $b^{*}(k)$ obeying CCR
\alpheqn
\eq
\ccr{b(k)}{b^*(p)} = \ddel(k-q)\quad\mbox{etc.}
\eqend
and a vacuum\footnote{Note that the term ``vacuum'' here and in the
following does {\em not} mean that
this state has anything to do with the ground state of the model; it is
just one convenient state from which all other states in the Hilbert space
can be generated by applying the field operators.} $\Omega_{\rm P}$ such that
\eq
b(k)\Omega_{\rm P} = 0 \quad \forall k\in\dLam .
\eqend
\reseteqn
We then set
\alpheqn
\eq
\label{photon}
\hat A_1(k) = \f{1}{s}\left(b(k) + b^*(k)\right)
\quad \hat E(k) = -\f{\ii s}{2}\left(b(k)-b^*(k)\right)
\eqend\reseteqn
where
$
s^4 = \pi e^2
$
(the reason for choosing this factor $s$ will become clear later).  We
will use below normal ordering $\xx\cdots\xx$ of bilinears in the Photon
field operators with respect to the vacuum $\Omega_{\rm P}$, for example
$\xx b(k)b^*(p)\xx\, = b^*(p)b(k)$.

{}For $\cH_{\rm Fermion}$ we take the Fermion Fock space with vacuum
$\Omega_{\rm F}$ such that
\eqa
\label{11}
\f{1}{2}(1 \pm \gamma_5) \hat\psi(\pm q) \,\Omega_{\rm F} &=&
\f{1}{2}(1 \mp \gamma_5) \hat\psi^*(\mp q) \,\Omega_{\rm F} =0
\quad \forall q > 0\, .
\eqaend
The presence of the Dirac sea requires
normal-ordering $\normal{\cdots}$ of the Fermion bilinears such as
$\hat H_0=\dIntF\Del q \normal{q\,\hat\psi^*(q)\gamma_5\hat
\psi(q)}$ and $\hat \rho_\pm$ (\ref{chiralcu}).
This modifies their naive commutator relations following
from the CAR as Schwinger terms show up \cite{GLrev,CR,A}.
In our case, the relevant commutators are:
\eqa
\ccr{\hat\rho^\pm(k)}{\hat\rho^{\pm}(p)}&=&
\pm k\ddel(k+p)\,, \nonumber\\
\ccr{\hat\rho^\pm(k)}{\hat\rho^{\mp}(p)} &=& 0  \label{12} \, , \\
\ccr{\hat H_0}{\hat\rho^\pm(k)} &=& \pm k
\hat\rho^\pm(k) \nonumber \, .
\eqaend
We note that
\eq
\label{vacF}
\hat\rho^+(k)\Om_{{\rm F}} = \hat\rho^-(-k)\Om_{{\rm F}} =
0\quad\forall k>0
\eqend
which together with \Ref{12} shows that the $\hat\rho^+(k)$ (resp.
$\hat\rho^-(k)$) give a highest (resp. lowest) weight representation of
the Heisenberg algebra.

We can now write the Gauss' law operators in Fourier space as
\eq
\label{14}
\hat G(k) = -\ii k\hat E(k) + e\hat\rho(k),
\eqend
so eqs. \Ref{12} imply
\[
\ccr{\hat G(k)}{\hat\rho^\pm(p)} =
\pm k e\ddel(k+p).
\]
Due to the presence of the Schwinger terms, these Fermion
currents no longer commute with the
Gauss' law generators, hence they are not gauge invariant
and no observables of the model.

To obtain Fermion currents obeying the classical relations (without
Schwinger terms), we note that $
\ccr{\hat G(k)}{\hat A_1(p)} =
k\ddel(k+p),$
hence the operators
\eq
\label{16}
\tilde\rho^\pm(k) \equiv \hat\rho^\pm(k)
 \pm e\hat A_1(k)
\eqend
commute with the Gauss law generators and are thus the observables of the
model corresponding to the chiral Fermion currents on the quantum
level.
Recalling the normalization is only unique up to finite terms, it is
natural to regard the $\tilde \rho^\pm(k)$ as the fermion
currents obtained by a {\em gauge covariant normal ordering}
preserving the classical transformation properties under gauge
transformations. Indeed, these currents can be shown to be
identical to those obtained by the gauge invariant point splitting method.

Similarly, the naive Hamiltonian $\hat H=\hat H_1+\hat H_2$,
\nonueqa
\hat H_1 &=& \hat H_0 +\dInt\Del k \xx \left(
\f{1}{4\pi}\hat E(k)\hat E(-k) + e\hat A_1(k)\hat j(-k)\right)\xx \\
\hat H_2 &=& \dInt\Del k \, \hat\rho^+(k)W_k \hat\rho^-(-k)
\nonueqaend
is not gauge invariant: $\hat H_1$ -- which is the naive Hamiltonian of the
Schwinger model -- obeys
\[
\ccr{\hat G(k)}{\hat H_1} = 2ke^2 \hat A_1(k)
\]
and therefore becomes gauge invariant only after adding a photon mass
term \cite{Manton}
\[
\dInt\Del k \, e^2 \hat A_1(k)\hat A_1(-k)
\]
(note that in position space this mass term has the usual form
$\f{e^2}{2\pi}\Int\dd{x}A_1(x)^2$, i.e.\ the photon mass--squared is
$e^2/\pi$).  Also the Luttinger--interaction term $\hat H_2$ 
becomes  gauge invariant only if one replaces the non--gauge
invariant currents $\hat\rho^\pm$ by the gauge invariant $\tilde\rho^\pm$ 
ones.

Thus we obtain the gauge invariant Hamiltonian of the
Luttinger--Schwinger model as follows,
\eqa
H= \hat H_0 + \dInt\Del k \xx\left( \f{1}{4\pi}
\hat E(k) \hat E(-k) + e\hat A_1(k)
\hat j(-k) + e^2 \hat A_1(k)\hat A_1(-k)
\right.\nonu\left.\f{}{}\!\!\!+
 \left[\hat\rho^+(k)+e\hat A_1(k)\right]
 W_k\left[\hat\rho^-(-k)-e\hat A_1(-k)\right]\right)\xx.
\eqaend

We can now explain the choice (\ref{photon},b) for the representation of the
Photon field: the factor $s$ is determined such that the free Photon
Hamiltonian
is equal to $\dInt\Del k \sqrt{\f{e^2}{\pi}}\, b^*(k)b(k)$.

\subsection{Bosonization}

Kronig's identity\footnote{in the modern literature this is often referred
to as (special case of the) Sugawara construction} allows us to rewrite the
free Hamiltonian as
$
\hat H_0 = \f{1}{2}\dInt\Del k \, \xx \left(
\hat\rho^+(k)\hat\rho^+(-k) \right.$ $\left.+ \hat\rho^-(k)
\hat\rho^-(-k) \right) \xx
$
(cf.\ Appendix A for the precise definition of normal ordering;
for simplicity of notation we do not distinguish the normal
ordering symbol for the photon fields and the fermion currents).  With
that, it follows from eq.\ \Ref{16} that
\eq
\label{20}
H = \dInt\Del k\, \xx \left(\f{1}{2}\left(
\tilde\rho^+(k)\tilde\rho^+(-k) + \tilde\rho^-(k)
\tilde\rho^-(-k) \right) + \f{1}{4\pi}\hat E(k) \hat E(-k) +
\tilde\rho^+(k) W_k \tilde\rho^-(-k)\right)\xx
\eqend
which is now explicitly gauge invariant.

\section{Solution of the model}

\subsection{Gauge Fixing}
The only gauge invariant degree of freedom of the Photon field at fixed
time is the holonomy $\int_\Lam\dd{x} A_1(x)$ and one can gauge away
all Fourier modes $\hat A_1(k)$ of the gauge field except the one for
$k=0$. Thus we can impose the gauge condition
\eq
\label{22}
\hat A_1(k) = \delta_{k,0} Y \makebox{, \qquad } A_1(x)=\f{2\pi}{L}Y
\eqend
and solve the Gauss' law $\hat G(k)\simeq 0$ (cf.\ eq.\ \Ref{14}) as
\eq
\hat E(k) \simeq
\f{e\hat\rho(k)}{\ii k} \quad {\mbox{for $k\neq 0$}}.
\eqend

This determines all components of $E$ except those conjugate to
$Y$: $\hat E(0) = \f{L}{2\pi} \f{\partial}{\ii \partial Y}$. After
that we are left with the ($k=0$)--component of Gauss' law, {\em viz.}
\eq
eQ_0\simeq 0,\quad Q_0 = \hat\rho(0) = \hat\rho^+ (0) + \hat\rho^- (0)\, .
\eqend
Inserting this into \Ref{20}, gives the Hamiltonian of the model on the physical
Hilbert space $\cH_{\rm phys}={\cal L}^2(\R,{\rm d}Y)\otimes {\cal H}'_{\rm 
Fermion}$
(where ${\cal H}'_{\rm Fermion}$ is the zero charge sector of the
fermionic Fock space):
\eqa
\label{24}
H = -\f{L}{8\pi^2} \f{\partial^2}{\partial Y^2} +
\f{\pi}{L}\left(\left(\hat\rho^+(0) + eY\right)^2 +
\left(\hat\rho^-(0) - eY\right)^2 + \left(\hat\rho^+(0) +
eY\right) 2 W_0 \left(\hat\rho^-(0) - eY\right)  \right) +
\nonu
\dIntp\Del k\, \xx \left(  \f{e^2}{4\pi k^2} \hat\rho(k)\hat\rho(-k)
 + \f{1}{2}\left( \hat\rho^+(-k)\hat\rho^+(k) +
\hat\rho^-(k)\hat\rho^-(-k) \right) +
 \hat\rho^+(k)W_k \hat\rho^-(-k)\right)\xx.
\eqaend

\subsection{Diagonalization of the Hamiltonian}
\label{zeromode}
{}Following \cite{HSU} we now write
\eq
H = \f{2\pi}{L}\sum_{k\geq 0} h_k.
\eqend
Introducing boson creation-- and annihilation operators
\alpheqn
\eq
\label{crho}
c(k) = \left\{\bma{cc} \f{1}{\sqrt{|k|}}\hat\rho^+(k) & \mbox{ for
$k>0$}\\
\f{1}{\sqrt{|k|}}\hat\rho^-(k) & \mbox{ for $k<0$} \ema\right.
\eqend
obeying usual CCR
\eq
\ccr{c(k)}{c^*(p)} = \ddel(k-p)\quad \mbox{etc.}.
\eqend
We then get for $h_{k>0}$
\eq
h_k =\left(k + \f{e^2}{2\pi k}\right)\left(c^*(k)c(k)+c^*(-k)c(-k)
\right) + \left( kW_k +
\f{e^2}{2\pi k}\right)\left(c^*(k)c^*(-k)+c(k)c(-k) \right). \eqend
\reseteqn
{}For $k=0$ we introduce the quantum mechanical variables
\alpheqn
\eqa
\label{qmv}
P &=& \left( \hat\rho^+(0) -\hat\rho^-(0) + 2eY  \right)\, , \nonu
X &=& \ii\f{L}{2\pi}\f{1}{2e} \f{\partial}{\partial Y}
\eqaend
obeying Heisenberg relations, $\ccr{P}{X}=-\ii L/2\pi$,
which allow us
to write $h_0$ as Hamiltonian of a harmonic oscillator,
\eq
\label{zeromom0}
h_0= \f{e^2}{\pi} X^2 + \f{1}{4}(1-W_0)\, P^2 + \f{1}{4} (1+W_0)\, Q_0^2
-\f{1}{2}\sqrt{\f{e^2}{\pi}}\f{L}{2\pi}
\eqend
\reseteqn
(the last term stems from normal ordering $\,\xx\cdots\xx\,$ 
and is irrelevant for the following).

We can now solve the model by diagonalizing its decoupled
Fourier modes $h_k$ separately, with the help of
a boson Bogoliubov transformation preserving the CCR,
\eq
\label{BT}
C(k) = \cosh(\lambda_k) c(k) + \sinh(\lambda_k) c^*(-k)
\eqend
where $\lambda_k=\lambda_{-k}$. This leads to 
\alpheqn
\eq
h_k = \om_k\left(C^*(k)C(k) + C^*(-k)C(-k)\right) -2\eta_k
\f{L}{2\pi}
\eqend
if we choose
\eq
\tanh(2\lambda_k) = \f{2\pi k^2 W_k + e^2}{2\pi k^2  + e^2} \label{th}\, .
\eqend
Then
\eq
\om_k^2 = k^2(1-W_k^2) + \f{e^2}{\pi}(1-W_k)
\eqend
and
\eq
\eta_k = \f{1}{2}\left(|k| + \f{e^2}{2\pi|k|} - \om_k \right)
\quad (k\neq 0) .
\eqend
\reseteqn

The zero--momentum piece $h_0$ is just
a harmonic oscillator and can be written as
\alpheqn
\eq
\label{zeromom}
h_0 = \om_0 C^*(0) C(0) + \f{1}{4} (1+W_0)\, Q_0^2 - \eta_0\f{L}{2\pi}
\eqend
with
\eq
\label{Cnull}
C(0) = \f{1}{\sqrt 2}\left(rX + \f{1}{r}\ii P\right),\quad
r^4=\f{e^2}{\pi}\f{4}{1-W_0},
\eqend
energy--squared
\eq
\om_0^2 = \f{e^2}{\pi}(1- W_0)
\eqend
and zero point energy
\eq
\eta_0 =\f{1}{2}\left(\sqrt{\f{e^2}{\pi}} -
\sqrt{\f{e^2}{\pi}(1-W_0)}\, \right).
\eqend
\reseteqn
Thus we get the Hamiltonian in the following form
\alpheqn
\eq
H=\dInt\Del k  \om_k C^*(k)C(k) -L E_0
\eqend
with the ground state energy density given by
\eq
E_0=\f{1}{2\pi}\dInt\Del k\, \eta_k.
\eqend
\reseteqn
(Note that for large $|k|$,
$
\eta_k = \f{1}{2}\left( \f{1}{2}|k W_k^2| + \f{e^2}{2\pi
|k|}W_k\right)\left(1+\OO\left(\f{1}{|k|}\right)\right),
$
hence $E_0$ is finite due to our assumptions \Ref{condition1} on
the potential.)

We now construct the unitary operator $\cU$ implementing the
Bogoliubov transformation \Ref{BT}, i.e.\
\eq
C(k) = \cU c(k)\cU^* \quad \forall k\in\dLam .
\eqend It is easy to see that operators $\cU_k$ satisfying $C(\pm k)
= \cU_k
c(\pm k)\cU_k^*$ for all $k>0$ are given by
\alpheqn
\eq
\label{cU}
\cU_k = \ee{S_k},\quad S_k = \lam_k \left( c(k)c(-k) - c^*(k)c^*(-k)
\right) \eqend  which are unitary since the operators $S_k$ are
screw-hermitian.\footnote{i.e.\ $\ii S_k$ is selfadjoint}
Thus,
\eq
\cU = \ee{S},\quad S=\sum_{k>0}S_k .
\eqend This operator $S$ can be shown to exist and defines an
anti-selfadjoint operator if and only if
\eq
\sum_{k>0}|k||\lam_k|^2<\infty \label{24c},
\eqend\reseteqn
and therefore (\ref{24c}) is necessary and sufficient for the unitary
operator $\cU$ to exist. This latter condition is equivalent to the
second one in \Ref{condition1} and thus fulfilled by assumption. Note
that
\eq
\label{hdiag}
\cU^* H \cU = \f{2\pi}{L}h_0 + \f{2\pi}{L}\sum_{k>0} \left(
\om_k\left( c^*(k)c(k) + c^*(-k)c(-k)\right)  -2\eta_k\f{L}{2\pi}\right)
\equiv H_D
\eqend
and therefore $\cU$ is the unitary operator diagonalizing the
non--zero modes of our Hamiltonian.

\subsection{Gauge invariant states}
\label{GND}
By the gauge fixing above we reduced the Hilbert space from $\cH$ to
$\cH'_{{\rm phys}}$ containing all states invariant under {\em small} gauge
transformations, i.e.  of the form $\ee{\ii\alpha(x)}$ with
$\alpha(L/2)=\alpha(-L/2)$.  There are, however, still large gauge
transformations present which are generated by $\ee{\ii2\pi x/L}$.  It is
important to note that physical states need not be invariant under these
latter transformations, but it is useful to construct states with simple
transformation properties.  This is the origin of the $\tet$--vacuum.

The large gauge transformation $\ee{\ii2\pi x/L}$ acts on the fields 
as follows
\eqa
\label{large}
\psi(x)&\tto& \ee{\ii 2\pi x/L}\psi(x) = (R_+R_-)^{-1}
\psi(x)(R_+R_-)\, ,\nonu
eY&\tto& eY - 1
\eqaend
where $R_\pm$ are the implementers of 
$\ee{\ii 2\pi x/L}$ in the chiral sectors of the
fermions and are discussed in detail
in Appendix A. The large gauge transformation $R$ obviously generates
a group $\Z$,  $n\to R^n$,  and we denote this group as $\Z_R$.
Our aim is to construct the states in
$\cH_{{\rm phys}}$ which carry an irreducible representation of
$\Z_R$ and especially the ground states of our model.

We start with recalling that the Fermion Fock space can be decomposed in
sectors of different chiral charges $\hat\rho^\pm(0)$,
$$\cH_{{\rm Fermion}} =\bigoplus_{n_+,n_-\in\Z}\cH^{(n_+,n_-)}$$
where
$$
\cH^{(n_+,n_-)}=\left\{\left. \Psi\in \cH_{{\rm Fermion}}\right|
\hat\rho^\pm(0)\Psi = n_\pm \Psi \right\} =
R_+^{n_+}R_-^{-n_-}\cH^{(0,0)}
$$
(for a more detailed discussion see Appendix A).
Thus,
\eq
\cH_{{\rm phys}}= {\cal L}^2(\R,\dd Y)\otimes \cH_{{\rm Fermion}}'
\eqend
where
\eq
\cH_{{\rm Fermion}}' = \bigoplus_{n\in\Z}\cH^{(n,-n)},
\quad \cH^{(n,-n)} = (R_+ R_-)^n\cH^{(0,0)}
\eqend
is the zero charge subspace of the Fermion Fock space and we
use the Schr\"odinger representation for the physical
degree of freedom $Y = \Int \dd{x} A_1(x)/2\pi$ of the photon
field as discussed in the last subsection.
$\cH_{{\rm phys}}$ can therefore be spanned by
states
\alpheqn
\eq
\label{tetn}
\Psi(n) = \phi\mbox{$(Y+\f{n}{e}) $} (R_+R_-)^n
\Psi,\quad \phi \in {\cal L}^2(\R,\dd Y),\, \Psi\in\cH^{(0,0)}
\eqend
which, under a large gauge transformation \Ref{large}, transform as
\eq
\Psi(n)\tto  \Psi(n-1).
\eqend
\reseteqn
Thus the states transforming under an irreducible representation of $\Z_R$
are given by
\eq
\label{tet}
\Psi^\tet= \sum_{n\in\Z}\ee{\ii\tet n}\Psi(n)
\tto \ee{\ii\tet}\Psi^\tet
\eqend
It is easy to calculate the inner products of these states,
\eq
<\Psi_1^\tet,\Psi_2^{\tet'}>=
2\pi\delta_{2\pi}(\tet-\tet')
<\Psi_1,\Psi_2>_{{\rm F}}(\phi_1,\phi_2)_{{\cal L}^2}
\eqend
($2\pi\delta_{2\pi}(\tet)=\sum_{n\in\Z}\ee{\ii n\tet}$,
since $<(R_+R_-)^n\Psi_1,(R_+R_-)^m\Psi_2> =
\delta_{n,m}<\Psi_1,\Psi_2>_{{\rm F}}$; $<\cdot,\cdot>_{{\rm F}}$ and
$<\cdot,\cdot>_{{\cal L}^2}$ are the inner products in $\cH_{{\rm Fermion}}$
and ${\cal L}^2(\R,\dd Y)$, respectively).
Thus the states $\Psi^\tet$ actually are not elements in
$\cH_{{\rm phys}}$ (they do not have a finite norm).

In our calculation of Green functions below we find it useful to use the
notation
\eq
\label{reg}
<\Psi_1^\tet,\Psi_2^{\tet}>_\tet \equiv
<\Psi_1,\Psi_2>_{{\rm F}}(\phi_1,\phi_2)_{{\cal L}^2}
\eqend
which can be regarded as redefinition of the inner product using a simple
multiplicative regularization (dropping the infinite term
$2\pi\delta_{2\pi}(0)$).

We now construct the ground states of our model. As expected, the
quantum mechanical variables $P,X$ \Ref{qmv} describing the zero mode $h_0$
of the Hamiltonian have a simple representation on the
$\tet$-states
\Ref{tet},
\eqa
P\Psi^\tet =  \sum_{n\in\Z}\ee{\ii\tet n}
2e\mbox{$(Y+\f{n}{e})$}
\phi\mbox{$(Y+\f{n}{e}) $}(R_+R_-)^n\Psi \, ,\nonu
X\Psi^\tet = \sum_{n\in\Z}\ee{\ii\tet n}
\f{\ii}{2e}\f{L}{2\pi}\f{\partial}{\partial Y}
\phi\mbox{$(Y+\f{n}{e}) $} (R_+R_-)^n\Psi.
\eqaend
Thus the ground states of $h_0$ annihilated by $C(0)$ are of the form
\Ref{tetn} with

\eq
\label{gst}
\phi_0(Y)
= \left(\f{\pi}{4e^2\al}\right)^\f{1}{4} \exp\left(-\alpha(2eY)^2\right)
\eqend
where
$
\label{alpha}
\alpha=\f{1}{L}\sqrt{\f{\pi^3}{2e^2}(1-W_0)},
$
and the other eigenstates are the harmonic oscillator eigenfunctions
$\phi_n\propto C^*(0)^n\phi_0$. From $C(k)=\cU c(k)\cU^*$ and
$c(k)\Om_{{\rm F}}=0$ it is clear that the ground state of all $h_{k>0}$
is $\cU \Om_{{\rm F}}$. We conclude that the ground states of our
model obeying $H \Psi_0^\tet = L E_0\Psi_0^\tet$ are given by
\eq
\label{vac}
\Psi_0^\tet= \sum_{n\in\Z}\ee{\ii\tet n}
\phi_0\mbox{$(Y+\f{n}{e}) $}(R_+R_-)^n \cU\Om_{{\rm F}}.
\eqend

\subsection{Gauge invariant Green functions}
\label{green}
The observables of our model now are operators on $\cH_{\rm phys}$ where
$\int_\Lam \dd x A_1(x)$ is represented by $2\pi Y$.
We recall that the fully gauge invariant field operators are the
$\chi$, \Ref{chi}, which are 
represented in the present gauge fixed setting by 
$$\chi_\sigma(x)=\ee{\ii 2\pi e Y(x-r)/L}\psi_\sigma(x).$$ 
These operators depend on the $r\in\Lam$ chosen.
Bilinears such as meson operators are, however, independent of $r$ and give
rise to translational invariant equal time Green functions. 
Moreover, on the quantum level not only the
Wilson line $W[A_1]$ \Ref{Wilson} but actually even
\eq
e \int_\Lam\dd x A_1(x) + \half Q_5  \equiv  w[A_1]
\eqend
is gauge invariant (note that $W[A_1] = \ee{\ii w[A_1]}$).
This operator is represented by $e Y+\half Q_5 = P/2$ (cf. \Ref{qmv}).

The gauge invariant equal time Green functions of the model are the ground
state expectation values of products $(\cdots)$ of meson operators and
functionals $F[P,X]$ of the zero mode operators $P$, $X$.  Since we only
consider $(\cdots)$ which are also invariant under large gauge
transformations, the transition amplitudes
$\left<\Psi_1^\tet,(\cdots)\Psi_2^{\tet'} \right>$ are always proportional
to $2\pi\delta(\tet-\tet')$.  Thus the Green functions we consider can be
defined as
\eq
\label{greenf}
\left< \Psi_0^\tet, F[P,X] \,
\chi^*_{\sigma_1}(x_1)\chi_{\tau_1}(y_1) \cdots
\chi^*_{\sigma_N}(x_N) \chi_{\tau_N}(y_N)\Psi_0^\tet \right>_\tet
\eqend
(note that $\left< \Psi_0^\tet, \Psi_0^\tet \right>_\tet =1$,
cf. \Ref{reg}).

{}Following \cite{HSU} it is useful to define {\em interacting
fermion fields}
\alpheqn
\eq
\label{intf}
\Psi_\sigma(x) = \cU^* \psi_\sigma(x)\, \cU
\eqend
such that (\ref{greenf}) becomes
\eq
\mbox{Eq.\ \Ref{greenf}} =
\left<\Om^\tet, F[P,X]\,  \Psi^*_{\sigma_1}(x_1)
\Psi_{\tau_1}(y_1)\cdots \Psi^*_{\sigma_N}(x_N)
\Psi_{\tau_N}(y_N)  \Om^{\tet'}\right>
\eqend
where
\eq
\label{freevac}
\Om^\tet= \sum_{n\in\Z}\ee{\ii\tet n}
\phi_0\mbox{$(Y+\f{n}{e}) $}(R_+R_-)^n \Om_{{\rm F}}
\eqend
\reseteqn
is the $\tet$--state constructed from the free fermion vacuum.

The strategy to calculate Green functions of the model using bosonization
techniques is the following: the relation \Ref{k2} of appendix A 
can be used to move
the operators $R_\pm$ and
combine them to some power of $(R_+R_-)$.  The operators $Q_\pm$
when applied to physical states become simple $\C$--numbers:
$Q_\pm (R_+R_-)^n = (R_+R_-)^n (\pm n+Q_\pm)$ for all integers
$n$, and $Q_\pm\Omega_{\rm F} = 0$. For the exponentials of boson operators
we use the decomposition into creation and annihilation parts
outlined in A.4. The normal ordering procedure gives a 
product of exponentials of commutators which are ($\C$-number) functions.
For the correlation functions of meson operators
$\chi_\si^*(x)\chi_{\si'}(y)$ we obtain:
\alpheqn
\eqa
\left<\Psi_0^\tet,\chi^*_{\pm}(x)
\chi_{\pm}(y)\Psi_0^{\tet} \right>_\tet &=&
\ee{-\f{\pi}{4L}m(x-y)^2} \ee{\De(x-y)}g_0^\pm(x-y) \label{2p1}\, , \\
\left<\Psi_0^\tet,\chi_{\pm}^*(x)
\chi_{\mp}(y)\Psi_0^{\tet} \right>_\tet &=& \ee{\mp \ii\tet}
\ee{-i\f{2\pi}{L}(x-y)} \ee{-\f{\pi m}{4L}((x-y)+\f{2}{m})^2} C(L) \ee{D(x-y)}
\label{2p2}
\, .
\eqaend
\reseteqn
with
\eqa
\Delta &=&
\sum_{k>0}\frac{2\pi}{Lk}\sinh^2(\la_k)[\ee{\ii kx }+
\ee{-\ii kx }-2] \nonumber\, , \\
D(x) &=& -\sum_{k>0} \frac{\pi}{Lk} \sinh(2\la_k)
[ \ee{\ii  kx }+ \ee{-\ii kx }-2] \label{DeDC}\, , \\
C(L) &=& \f{1}{L}\exp[\sum_{k>0} \f{2\pi}{kL}(\sinh(2\la_k)-2\sinh^2(\la_k))]\,.
\nonumber
\eqaend
where $g_0^\pm(x)=\f{1}{L}\f{e^{\mp i\f{\pi}{L}x}}{1-e^{\pm i\f{2\pi}{L}
(x\pm i\eps)}}$
is the 2-point function of free fermions, and the Schwinger mass is
renormalized to $m^2=\f{e^2}{\pi(1-W_0)}$.

Note that the Green function \Ref{2p2}
depends on $\tet$ and is non--zero due to chiral symmetry breaking as in
the Schwinger model.  As expected, for vanishing electromagnetic coupling,
$e=0$, this Green function vanishes (due to the factor $\ee{-\pi/mL}$ appearing
in (\ref{2p2})).

{}From  (\ref{2p2}) we can calculate the chiral condensate by
setting $x=y$, and in the  limit $L\to\infty$ we obtain
\eq
\lim_{L\to\infty}
\left<\Psi_0^\tet,\chi_{\pm}^*(x) \chi_{\mp}(x)\Psi_0^{\tet} \right>_\tet
= \lim_{L\to\infty} \ee{\mp \ii\tet} \ee{-\f{\pi}{mL}}
C(L) \nonumber
= \ee{\mp \ii\tet} C
\eqend
with a constant $C$ which can be calculated in principle
from eq.\ (\ref{DeDC}).  In
the special case of the Schwinger model ($W_k=0$), $C$ can be computed
and we recover the well--known result $C_{W_k=0}=\f{m}{4\pi}\ee{\ga}$ where
$\ga=0.577\ldots$ is Eulers constant
(see e.g.\ \cite{SaWi,Hoso}).

\section{Multiplicative regularization and the Thirring-Schwinger model}

We recall that the Thirring model is formally obtained
from the Luttinger model in the limits
\eq
\label{LT}
 L\to \infty, \quad V(x)\to g \delta(x)
\eqend
i.e.\ when the interaction becomes local and space becomes infinite.
The first
limit amounts to remove the IR cut--off of our
model.  By inspection it can be easily done in all Green functions.
The second limit in \Ref{LT} is non--trivial: we recall, that condition
\Ref{condition1} on the Luttinger potential requires
sufficient decay of the Fourier modes $W_k$ of the interaction, and this is
violated in the Thirring model where $W_k=W_0$ is independent of $k$.  This
latter condition was necessary for the interacting model to be
well--defined on the Hilbert space of the non--interacting model.

A better understanding can be obtained by explicitly
performing the limit \Ref{LT} in the present setting.  
The idea is to find a family of Luttinger
potentials $\{ V_\ell(x)\}_{\ell>0}$ becoming local for $\ell\downarrow 0$,
i.e.\ for all $\ell>0$ the condition \Ref{condition1} is fulfilled and
$\lim_{\ell\downarrow 0} V_\ell(x)=g\delta(x)$.  Then for all $\ell>0$
everything is well-defined on the free Hilbert space and one can work out in
detail how to regularize such that the correlation functions make sense
for $\ell\downarrow 0$.

We note that a direct construction of the Thirring model in a
framework similar to the one here has been completed in \cite{CRW}. This
construction seems to be, however, different from the one outlined below.

{}For the case of Luttinger-Schwinger model we split the function $\De(x)$ 
into a part corresponding to the pure Luttinger model and a part which 
describes the additional Schwinger coupling, 
i.~e.~$\De(x)=(\De(x)-\De^{e=0}(x))+\De^{e=0}(x)$.

The limit $W_k=W_0={\rm const.}$ exists for $\De-\De^{e=0}$. As $L\to\infty$, the sum
in (\ref{DeDC}) turns into an integral and we obtain 
\eqa
\De(x)-\De^{e=0}(x)=&&\!\!\!\!\!\!\! \f{1}{\sqrt{1-W_0^2}} \int_0^\infty {\rm d}k\, \left(
\f{1}{\sqrt{k^2+\mu^2}}-1\right)(\cos (kx) -1)+ \nonumber\\
&& \!\!\!\!\!\!\!
\sqrt{\f{1+W_0}{1-W_0}}\, \int_0^\infty {\rm d}k\, \f{\mu^2}{k^2}\f{1}{\sqrt{k^2+\mu^2}}
(\cos (kx) -1) \, .
\eqaend
The first integral becomes $K_0(|\mu x|)+\ln\f{|\mu x|}{2}+\ga$ and the
expression in the last line is a second integral $(n=2)$ 
of $K_0$ defined iteratively
by Ki$_n(x)=\int_x^\infty $Ki$_{n-1}(t) \, dt$, Ki$_0=K_0$ \cite{Abram}. 
Moreover we
introduced a new mass by $\mu^2=e^2/(\pi(1+W_0))$. Note that the
singularities at the origin of the Bessel function are removed by the 
additional terms, consistent with $\De(0)=0$.

No regularization has been necessary so far. Renormalization comes along with
$\De^{e=0}$. 
We choose a Luttinger-interaction such that
$(1-W_k^2)^{-1/2}-1=2a^2 e^{-\ell k}$ where $\ell$ 
defines the range of the interaction.
For this choice  we find
\eqa
\De^{e=0}(x)=2a^2\ln\left|\f{\ell}{x+i\ell}\right|
\eqaend
and obviously the Thirring limit makes sense only if one removes the singular
part $\ln \ell$ which can be done by a wave function renormalization of the form
\eqa
\chi_\pm(x) \to \tilde\chi_\pm(x)=Z^{1/2}(a,\ell)\chi_\pm(x)
\quad \makebox{with}
\quad Z^{1/2}(a,\ell)=\ell^{-a^2} \, . \label{Th3}
\eqaend
A similar discussion holds for the chirality mixing correlation function.
The 2-point function of the Thirring-Schwinger model therefore
become
\alpheqn
\eqa
\langle\Psi_0^\tet,\tilde\chi_\pm^*(x)\tilde\chi_\pm(0)\Psi_0^\tet\rangle_\tet
&=&e^{\De_{\rm reg}(x)}g_0^\pm(x)\, ,\\
\left<\Psi_0^\tet,\tilde\chi_{\pm}^*(x)
\tilde\chi_{\mp}(0)\Psi_0^{\tet} \right>_\tet &=& \ee{\mp \ii\tet}
C_{\rm reg}\, \ee{D_{\rm reg}(x)}
\, .
\eqaend
\reseteqn
If we define $\tau_0$ by $\tanh(2\tau_0)=W_0$ we can write
\eqa
\De_{\rm reg}(x)&\!=&\!\cosh (2\tau_0)\left[ K_0(|\mu x|)+\ln\f{|\mu x|}{2}+\ga\right]
+\nonumber\\
&&\!\!\f{1}{2}e^{2\tau_0}\left[1-\f{\pi}{2}|\mu x|-{\rm Ki}_2(|\mu x|)\right]+
(\cosh (2\tau_0)-1)\ln |x| \, , \nonumber\\
D_{\rm reg}(x) &\!=&\! -\sinh (2\tau_0)\left[ K_0(|\mu x|)+\ln\f{|\mu x|}{2}+\ga\right]
-  \\
&&\!\! \f{1}{2}e^{2\tau_0}\left[1-\f{\pi}{2}|\mu x|-{\rm Ki}_2(|\mu x|)
\right] \, ,\nonumber\\
\ln C_{\rm reg} &\!=&\! \ga+\ln\f{1}{2\pi}+e^{-2\tau_0}\ln \f{\mu}{2} 
\nonumber \, .
\eqaend
We checked that all Green functions of the Thirring-Schwinger model have
a well-defined limit after the wave function renormalization.

We would like to stress that this procedure can be naturally interpreted as
low--energy limit of the Luttinger--Schwinger model: 
if one is interested only in
Green functions describing correlations of far--apart fermions, the precise
form of the Luttinger interaction $V(x)$ should be irrelevant and only the
total interaction strength $g = \int\dd x\, V(x)$ should matter.  Thus as
far
as these correlators are concerned, they should be equal to the ones of the
Thirring model corresponding to this coupling $g$.

\section{Conclusion}
We formulated and solved the Luttinger-Schwinger model in the Hamiltonian
formalism.  Structural issues like gauge invariance, the role of anomalies
and the structure of the physical states were discussed in detail.  The
necessary tools for computing all equal time correlation functions were
prepared and illustrated by calculating the 2--point Green functions.  From
this the chiral condensate and critical exponents were computed.  We could
also clarify how the non trivial short distance behavior of the
Thirring-Schwinger
model arises in a limit from the Luttinger-Schwinger model.


\app
\section*{Appendix A: Bosons from fermions and vice versa}

In this appendix we summarize the basics for the bosonization used in the
main text to solve the Luttinger--Schwinger model.  Bosonization is known
in the physics literature since quite some time 
(\cite{CR,PressSegal,Kac,Mickelsson}), for a discussion of the
older history see \cite{HSU}). 

We consider the fermion Fock space $\cH_{\rm Fermion}$ generated
by the fermion field operators
from the vacuum $\Om_{\rm F}$ as described in
the main text.  We note that $\cH_{\rm Fermion}=\cH_{+}\otimes\cH_{-}$
where $\cH_{\pm}$ are generated by the left-- and right--handed chiral
components $\hat\psi_+$ and $\hat\psi_-$ of our Dirac fermions.
Bosonization can be 
formulated for the chiral components $\hat\psi_\pm$ separately
as it leaves
the two chiral sectors $\cH_{\pm}$ completely decoupled.  For our purpose
it is more convenient to treat both chiral sectors together.

\subsection*{A.1 Structure of fermion Fock space}
We start by introducing two unitary operators $R_\pm$
which are defined up to an irrelevant phase factor (which we will leave
unspecified) by the following equations,
\eq
\hat\psi_\pm(k)R_\pm  = R_\pm \hat\psi_\pm (k - \f{2\pi}{L})
\eqend
and $R_\pm$ commutes with $\hat\psi_\mp$. A proof of existence and an
explicit construction of these operators can be found in \cite{Ruij}.
Here we just summarize their physical meaning and special properties.

It is easy to see that $R_\pm$ are just the implementors of Bogoliubov
transformations given by the {\em large gauge transformations}
$\psi_\pm(x)\mapsto \ee{\ii 2\pi x/L}\psi_\pm(x)$
and $\psi_\mp(x)\mapsto \psi_\mp(x)$, hence $R_+R_-$ and
$R_+R_-^{-1}$ implement
the vector-- and  the axial large gauge transformations
$\ee{\ii 2\pi x/L}$ and $\ee{\ii\gamma_5 2\pi  x/L }$, respectively.
These have non--trivial winding number\footnote{
the w.n. of a smooth gauge transformation $\Lam\to{\rm U}(1)$,
$x\mapsto \ee{\ii\alpha(x)}$ is the integer
$\f{1}{2\pi}\left(\alpha(L)-\alpha(0)\right)$.
}
and change the vacuum $\Om_{\rm F}$ to states containing (anti-)
particles.
The latter follows from the commutator relations with the chiral
fermion currents 
\eqa
(R_\pm)^{-1} \hat\rho^\pm (k) R_\pm = \hat\rho^\pm (k) \pm
\delta_{k,0}  \label{k2}\, .
\eqaend

The essential point of bosonization is that the total Hilbert space
$\cH_{\rm Fermion}$ can be generated from $\Om_{\rm F}$ by the chiral
fermion currents $\hat\rho^{\pm}(k)$ and $R_\pm$.  More
precisely, for all pairs of integers $n_+,n_-\in\Z$ we introduce the
subspaces $\cD^{(n_+,n_-)}$ of $\cH_{{\rm Fermion}}$ containing all linear
combinations of vectors
\eq
\label{lin}
\hat\rho^+(k_1) \cdots \hat\rho^+(k_{m_+}) \hat\rho^-(q_1)\cdots
\hat\rho^-(k_{m_-} ) R_+^{n_+} R_-^{-n_-} \Om_F
\eqend
where $m_\pm\in\N_0$ and $k_i,q_i\in\dLam$.
The basic result of the boson--fermion correspondence is the
following

{\bf Lemma:} The space
\eq
\cD \equiv \bigoplus_{n_+,n_-\in\Z} D^{(n_+,n_-)} .
\eqend
is dense in $\cH_{\rm Fermion}$ (for a proof see e.g.\ \cite{CR}).

{\em Remark:} This Lemma gives the following picture of the
structure of the Fock space $\cH_{\rm Fermion}$: It splits into
{\em superselection sectors} $\cH^{(n_+,n_-)}$ (which are the closure of
$\cD^{(n_+,n_-)}$) containing the eigenstates of the chiral charges
$Q_\pm$ with eigenvalues $n_\pm$. The fermion currents $\hat\rho^\pm(k)$
leave all these sectors invariant, and the operators $R_\pm$ intertwine
different sectors, $R_+: \cH^{(n_+,n_-)}\to \cH^{(n_+ +1,n_-)}$ and
$R_-: \cH^{(n_+,n_-)}\to \cH^{(n_+,n_- - 1)}$.

\subsection*{A.2 Kronig's identity}
The basic formula underlying the solution of our model is
\eq
\label{kronig}
\hat H_0 = \f{\pi}{L}\left( Q_+^2 + Q_-^2 \right) +
\f{2\pi}{L} \sum_{k>0}\left( \hat\rho^+(-k)\hat\rho^+(k) +
\hat\rho^-(k) \hat\rho^-(-k)  \right) .
\eqend
It expresses the free Dirac Hamiltonian in terms of bilinears of the
chiral fermion currents. 
\subsection*{A.3 Boson--fermion correspondence}
The boson--fermion correspondence provides explicit formulas of the
fermion operators $\psi_\pm(x)$ in terms of operators
$\hat\rho^\pm(k)$ and $R_\pm$,
\aalpheqn
\eq
\label{limit}
\psi_\pm(x) = \lim_{\eps\searrow 0} \psi_\pm(x;\eps)
\eqend
(this limit can e.g. be understood in the weak sense for states in
$\cD$), where
\eqa
\psi_\pm(x;\eps) = \f{1}{\sqrt{L}}
S_\pm(x) \normal{\exp(K_\pm(x;\eps))} \label{bfc}
\eqaend
with
\eq
\label{S}
S_\pm(x)=
\ee{\pm \ii \pi x  Q_\pm/L } (R_\pm)^{\mp 1}\ee{\pm \ii \pi x  Q_\pm/L } =
\ee{\mp \pi\ii x/L} (R_\pm)^{\mp 1}\ee{\pm \ii 2\pi x  Q_\pm/L }
\eqend
and
\eq
\label{Kpm}
K_\pm(x;\eps) =\mp \f{2\pi}{L} \sum_{k\in\dLam\backslash\{0\}}
\f{\hat\rho^\pm(-k)}{k}\ee{-\ii kx}\ee{-\eps|k|} = -
K_\pm(x;\eps)^*.  \eqend
\areseteqn
More explicitly, the normal ordering $\normal{\cdots}$ is with respect
to the fermion vacuum $\Om_{{\rm F}}$ (cf.\ \Ref{vacF}),
\aalpheqn
\eq
\normal{\exp(K_\pm(x;\eps))} = \exp(K^{(-)}_\pm(x;\eps))
\exp(K^{(+)}_\pm(x;\eps))
\eqend
where
\eq
K^{(\sigma)}_\pm(x;\eps) = \sigma\f{2\pi}{L}\sum_{k>0}
\f{\hat\rho^\pm(\pm \sigma k)}{k}\ee{\pm\sigma \ii kx}\ee{-\eps|k|},
\quad \sigma=+,-
\eqend
\areseteqn
is such that $K_\pm = K_\pm^{(-)}+K_\pm^{(+)}$ and
$K_\pm^{(+)}\Om_{{\rm F}} = 0$ (cf.\ \Ref{vacF}).

\subsection*{A.4 Interacting fermions}
{}From the definition of the interacting fermion fields $\Psi(x)$ (\ref{intf})
and the 
representation of free fermions in terms of bosons, we are led to investigate
the interacting kernel $\tilde K_\pm(x)={\cal U}^* K_\pm(x) {\cal U}$:
$$
\tilde{K}_\pm(x) =\mp \f{2\pi}{L} \sum_{k\in\dLam\backslash\{0\}}
\f{1}{k} \left( \cosh(\lambda_k)\, \hat\rho^\pm(-k) -
\sinh(\lambda_k)\, \hat\rho^\mp(-k) \right) \ee{-\ii kx}\ee{-\eps|k|}\, .
$$
It is convenient to write
\newcommand{\Ks}{{K\! s}}
\newcommand{\Kc}{{K\! c}}
\newcommand{\Ksc}{{K\! s/c}}
\aalpheqn
\eq
\tilde{K}_\pm = \tilde{K}_\pm^{(+)} + \tilde K_\pm^{(-)},\quad
\tilde{K}_\pm^{(\sigma)} = Kc_\pm^{(\sigma)} -Ks_\mp^{(\sigma)}
\eqend
where the upper index refers to the creation-- ($\sigma=-$) and
annihilation-- ($\sigma=+$) parts of operators and
\eqa
\Kc^{(\si)}_\pm(x) &=& \si\sum_{k>0}\frac{2\pi}{Lk}\cosh(\la_k)
\hat\rho^\pm(\pm\si k)\ee{\mp\si \ii kx }\ee{-\eps k} \, , \nonu
\Ks^{(\si)}_\pm(x) &=& \si\sum_{k>0}\frac{2\pi}{Lk}\sinh(\la_k)
\hat\rho^\pm(\pm\si k)\ee{\mp\si \ii kx }\ee{-\eps k} \: .
\eqaend
\areseteqn
The nonzero commutators of these operators are
\eqa
\ccr{\Kc^{(+)}_\pm(x)}{\Kc^{(-)}_\pm(y)} &=&
- \sum_{k>0}\frac{2\pi}{Lk}\cosh^2(\la_k)
\ee{ \mp \ii k(x-y) }\ee{-2\eps k} \, , \nonu
\ccr{\Kc^{(+)}_\pm(x)}{\Ks^{(-)}_\pm(y)} &=&
- \sum_{k>0}\frac{\pi}{Lk}\sinh(2\la_k)
\ee{ \mp\ii k(x-y) }\ee{-2\eps k}\, ,  \\
\ccr{\Ks^{(+)}_\pm(x)}{\Ks^{(-)}_\pm(y)} &=&
- \sum_{k>0}\frac{2\pi}{Lk}\sinh^2(\la_k)
\ee{ \mp\ii k(x-y) }\ee{-2\eps k} \: .
\nonumber
\eqaend
We find the following normal ordered expression for the interacting fermions
\eq
\Psi_\pm(x) = \f{1}{\sqrt{L}} z S_\pm(x)
\normal{\ee{\tilde K_\pm(x) }}
\eqend
where $z = \ee{- \sum_{k>0}
\f{2\pi}{Lk}\sinh^2(\la_k) }$.

\appende

\bc{\bf Acknowledgments}\ec
E.L.  would like to thank the Erwin Schr\"odinger International Institute
in Vienna for hospitality where part of this work was done, and the
``\"Osterreichische Forschungsgemeinschaft'' for partial financial support
in May/June 1994 when this work was begun.  He would also like to thank
S.G. Rajeev and Mats Wallin for usefull discussions. The authors thank the 
referee for valuable suggestions concerning the presentation of their results.

\end{document}